\documentclass[twocolumn,showpacs,preprintnumbers,amsmath,amssymb,superscriptaddress]{revtex4}

\usepackage{graphicx}
\usepackage{dcolumn}
\usepackage{bm}

\begin{document}


\title{Quenching Spin Decoherence in Diamond through Spin Bath Polarization}
\author{Susumu Takahashi}
\email{susumu@iqcd.ucsb.edu}
\affiliation{Department of Physics and Center for Terahertz Science and Technology, University of California, Santa Barbara, California 93106}%

\author{Ronald Hanson}
\affiliation{Kavli Institute of Nanoscience, Delft University of Technology, P.O. Box 5046, 2600 GA Delft, The Netherlands}%
\affiliation{Department of Physics and Center for Spintronics and Quantum Computation, University of California, Santa Barbara, California 93106}%

\author{Johan van Tol}
\affiliation{National High Magnetic Field Laboratory, Florida State University, Tallahassee Florida  32310}%

\author{Mark S. Sherwin}%
\affiliation{Department of Physics and Center for Terahertz Science and Technology, University of California, Santa Barbara, California 93106}%

\author{David D. Awschalom}%
\affiliation{Department of Physics and Center for Spintronics and Quantum Computation, University of California, Santa Barbara, California 93106}%

\date{\today}

\begin{abstract}
We experimentally demonstrate that the decoherence of a spin by a
spin bath can be completely eliminated by fully polarizing the
spin bath. We use electron paramagnetic resonance at 240 gigahertz
and 8 Tesla to study the spin coherence time $T_2$ of
nitrogen-vacancy centers and nitrogen impurities in diamond from
room temperature down to 1.3 K. A sharp increase of $T_2$ is
observed below the Zeeman energy (11.5 K). The data are well
described by a suppression of the flip-flop induced spin bath
fluctuations due to thermal spin polarization. $T_2$ saturates at
$\sim 250~\mu s$ below 2 K, where the spin bath polarization is
99.4 $\%$.
\end{abstract}

\maketitle

%
%

Overcoming spin decoherence is critical to spintronics and
spin-based quantum information processing
devices~\cite{awschalom07, hanson07}. For spins in the solid
state, a coupling to a fluctuating spin bath is a major source of
the decoherence. Therefore, several recent theoretical and
experimental efforts have aimed at suppressing spin bath
fluctuations~\cite{khaetskii02, merkulov02, desousa03,
dobrovitski03, cucchietti05, taylor06, yao07}. One approach is to
bring the spin bath into a well-known quantum state that exhibits
little or no fluctuations~\cite{stepanenko06, klauser06}. A prime
example is the case of a fully polarized spin bath. The spin bath
fluctuations are fully eliminated when all spins are in the ground
state. In quantum dots, nuclear spin bath polarizations of up to
$60\%$ have been achieved~\cite{bracker05, baugh07}. However, a
polarization above $90\%$ is need to significantly increase the
spin coherence time~\cite{coish04}. Moreover, thermal polarization
of the nuclear spin bath is experimentally challenging due to the
small nuclear magnetic moment. Electron spin baths, however, may
be fully polarized thermally at a few degrees of Kelvin under an
applied magnetic field of 8 Tesla.

Here we investigate the relationship between the spin coherence of
Nitrogen-Vacancy (N-V) centers in diamond and the polarization of
the surrounding spin bath consisting of Nitrogen (N) electron
spins. N-V centers consist of a substitutional nitrogen atom
adjoining to a vacancy in the diamond lattice. The N-V center,
which has long spin coherence times at room
temperature~\cite{kennedy03, gaebel06}, is an excellent candidate
for quantum information processing applications as well as
conducting fundamental studies of interactions with nearby
electronic spins~\cite{gaebel06, hanson06prl, hanson08} and
nuclear spins ~\cite{childress06, dutt07}. In the case of type-Ib
diamond, as studied here, the coupling to a bath of N electron
spins is the main source of decoherence for an N-V center
spin~\cite{kennedy03, hanson06}. We have measured the spin
coherence time ($T_2$) and spin-lattice relaxation time ($T_1$) in
spin ensembles of N-V centers and single N impurity centers (P1
centers) using pulsed electron paramagnetic resonance (EPR)
spectroscopy at 240 GHz. By comparing the values of $T_1$ and
$T_2$ at different temperatures, we verify that the mechanism
determining $T_2$ is different from that of $T_1$. Next, we
investigate the temperature dependence of $T_2$.

At 240 GHz and 8.6 T where the Zeeman energy of the N centers
corresponds to $11.5$ K, the polarization of the N spin bath is
almost complete (99.4 $\%$) for temperatures below 2 K as shown in
Fig.~1(a). This extremely high polarization has a dramatic effect
on the spin bath fluctuations, and thereby on the coherence of the
N-V center spin. We find that $T_2$ of the N-V center spin is
nearly constant between room temperature and 20 K, but increases
by almost 2 orders of magnitude below the Zeeman energy to a
saturation value of $\sim 250$ $\mu$s at 2 K. The data shows
excellent agreement with a model based on spin flip-flop processes
in the spin bath. The observed saturation value suggests that when
the N spin bath is fully polarized, $T_2$ is limited by the
fluctuations in the $^{13}$C nuclear spin bath.

%
%
We studied a single crystal of high-temperature high-pressure
type-Ib diamond, which is commercially available from Sumitomo
electric industries. The density of N impurities is $10^{19}$ to
$10^{20}$ cm$^{-3}$. The sample was irradiated with 1.7 MeV
electrons with a dose of 5 $\times$ $10^{17}$ cm$^{-3}$ and
subsequently annealed at 900 $^{\circ}$C for 2 hours to increase
the N-V concentration~\cite{epstein05}.

Electronic spin Hamiltonians for the N-V ($H_{NV}$) and N centers
($H_{N}$) are,
\begin{eqnarray}
H_{NV} = D[(S^{NV}_z)^2-\frac{1}{3}S(S+1)]\nonumber\\ +
\mu_{B}g^{NV}{\bm S^{NV}}\cdot{\bm B_0} + A^{NV}{\bm
S^{NV}}\cdot{\bm I^{N}},
\end{eqnarray}
\begin{eqnarray}
H_{N} = \mu_{B}{\bm S^{N}}\cdot \stackrel{\leftrightarrow}{g^{N}}
\cdot{\bm B_0} + A^N{\bm S^{N}}\cdot{\bm I^{N}},
\end{eqnarray}
where $\mu_B$ is the Bohr magneton and $\bm B_0$ is the magnetic
field. ${\bm S^{NV}}$ and ${\bm S^{N}}$ are the electronic spin
operators for the N-V and N centers and ${\bm I^{N}}$ is the
nuclear spin operator for $^{14}N$ nuclear spins.
$g^{NV}=2.0028$~\cite{loubser78}, and
$\stackrel{\leftrightarrow}{g^{N}}$ is the slightly anisotropic
g-tensor of the N center. $D=2.87$ GHz is the zero-field splitting
due to the axial crystal field~\cite{loubser78}. Due to the
tetrahedral symmetry of diamond lattice, there are four possible
orientations of the defect principal axis of the $^{14}N$
hyperfine coupling of $A^{N}$ and $A^{NV}$. In the present case,
$A^{N} = 114$ MHz for the $\langle$111$\rangle$-orientation and
$A^{N} = 86$ MHz for the other three orientations~\cite{smith59}.
For the N-V center, $A^{NV} = 2.2$ MHz for the
$\langle$111$\rangle$-orientation~\cite{loubser78}. The nuclear
Zeeman energy and the hyperfine coupling between the N-V (N)
center and $^{13}$C and the nuclear Zeeman energy are not included
here. The energy states of the N-V and N centers are shown in
Fig.~1(b).
\begin{figure}
\includegraphics[width=80 mm] {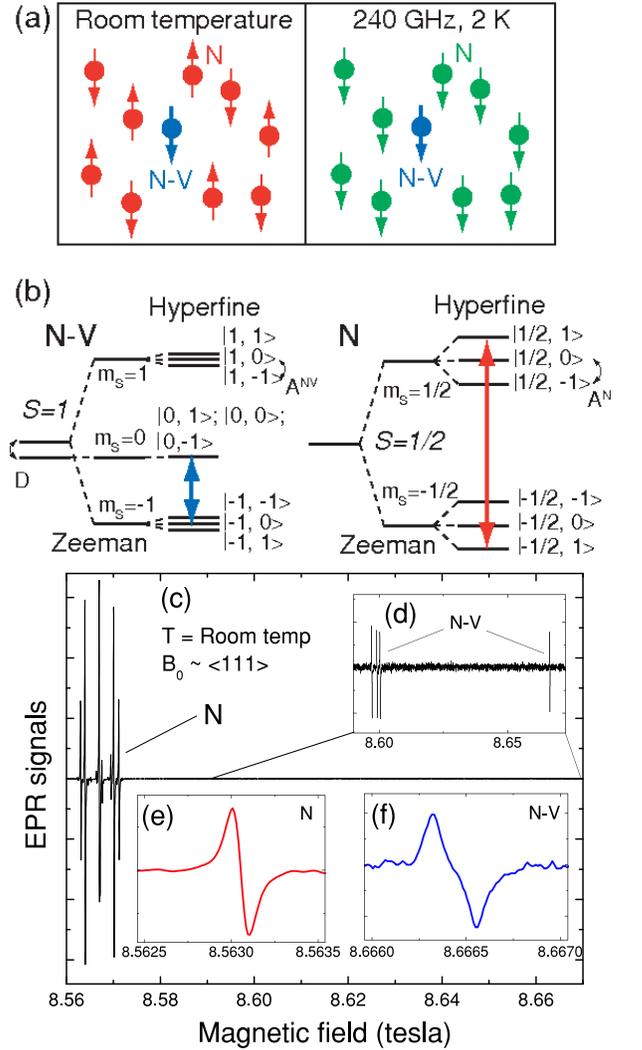}
\caption{\label{fig:cwEPR} (a)Spins of the N-V and N centers at
room temperature and at 8.56 tesla and 2 K. At room temperature,
where up and down spins are nearly equally populated, the N spin
bath polarization is very small and therefore, the spin flip-flop
rate is high. At 240 GHz and 2 K, the N spin bath polarization is
99.4 $\%$ and the spin flip-flop rate is nearly zero. (b)Energy
states of the N-V and N centers. The energy levels are not scaled.
The states are indexed by $\vert m_S, m_I\rangle$. Transitions
indicated by solid lines are EPR peaks used to measure the spin
relaxation times $T_1$ and $T_2$. (c)cw EPR spectrum at 240 GHz at
room temperature when the magnetic field $B_0$ is applied along
the $\langle$111$\rangle$-direction. No optical pump is applied.
The strongest five EPR peaks around 8.57 tesla are from N centers.
(d) N-V EPR peaks. The intensity ratio between the left-most N and
the right-most N-V is $\sim$ 80 which corresponds to 120:1
population ratio between N and N-V centers respectively. Other
impurity centers were also observed (not indicated). (e)N centers
EPR for the transition of $\vert m_S=-1/2, m_I=1 \rangle
\leftrightarrow \vert 1/2, 1\rangle$. (f)N-V centers EPR for the
transition of $\vert m_S=-1 \rangle \leftrightarrow \vert 0
\rangle$.}
\end{figure}

%
%
The measurement was performed using a 240 GHz continuous wave (cw)
and pulsed EPR spectrometer in the electron magnetic resonance
program at the National High Magnetic Field Laboratory (NHMFL),
Tallahassee FL. The setup is based on a superheterodyne
quasioptical bridge with a 40 mW solid state source. Details of
the EPR setup are described elsewhere~\cite{vantol05, morley08}.
No optical excitation was applied throughout this paper, and no
resonator was used for either cw or pulsed experiments.
Fig.~1(c)-(f) shows cw EPR spectra at room temperature where the
magnetic field was applied along the
$\langle$111$\rangle$-direction of the $\sim 0.8 \times 0.8 \times
0.6$ mm$^3$ single crystal diamond. The applied microwave power
and field modulation intensity were carefully tuned not to distort
the EPR lineshape. Five EPR spectra in Fig.~1(c) corresponding to
the N center are drastically stronger than the remaining signals
which indicates that the number of N centers dominates the spin
population in the sample. The N EPR peaks show the slightly
anisotropic g-factor $g^{N}$ which gives
$g^{N}_{\parallel}=2.0024$ and $g^{N}_{\perp}=2.0025 \sim 6$ and
is in agreement with the reported g-anisotropy of type-IIa
diamond~\cite{zhang94}. As shown in Fig.~1(d), we also observed
the much smaller N-V resonances which shows a line for the
$\langle$111$\rangle$-orientation in the right side and three
lines for the other orientations in the left side. An overlap of
the three lines is lifted because the applied $B_0$ field is
slightly tilted from the $\langle$111$\rangle$-direction. Based on
the EPR intensity ratio between N and N-V centers, the estimated
density of the N-V centers in the studied sample is approximately
$10^{17}$ to $10^{18}$ cm$^{-3}$. EPR lineshapes of the N ($\vert
m_S = -1/2, m_I=1\rangle \leftrightarrow \vert 1/2, 1\rangle$) and
N-V ($\vert m_S = -1 \rangle \leftrightarrow \vert 0 \rangle$)
centers are shown in Fig.~1(e) and (f) respectively. The N center
shows a single EPR line with a peak-to-peak width of 0.95 gauss.
On the other hand, the N-V center shows a broader EPR line (the
peak-to-peak width is 2.36 gauss) due to the hyperfine coupling
between the N-V center and the $^{14}$N nuclear spins. The
estimated hyperfine constant is 2 MHz, in good agreement with a
previous report~\cite{loubser78}.

%
%
The temperature dependence of the spin relaxation times $T_1$ and
$T_2$ was measured using pulsed EPR. An echo-detected inversion
recovery sequence ($\pi-T-\pi/2-\tau-\pi-\tau-echo$) is applied
for $T_1$ where a delay $T$ is varied, while a Hahn echo sequence
($\pi/2-\tau-\pi-\tau-echo$) is applied for $T_2$ where a delay
$\tau$ is varied~\cite{schweiger}. The area of the echo signal
decays as a function of the delay time $T$ and $2\tau$ for $T_1$
and $T_2$ respectively and therefore can be used to determine the
relaxation times. For the pulsed EPR measurement, we used the
$\vert m_S=-1, m_I=0\rangle \leftrightarrow \vert 0, 0\rangle$
transition for the N-V center and the $\vert m_S=-1/2, m_I=1
\rangle \leftrightarrow \vert 1/2, 1\rangle$ transition for the N
center (Fig.~1(b)).

\begin{figure}
\includegraphics[width=80mm]{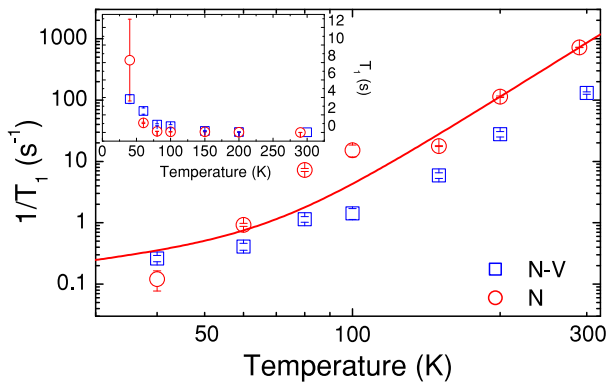}
\caption{\label{fig:invT1} $1/T_1$ for the N-V and N centers as a
function of temperature. Solid lines are the best fit of the
spin-orbit phonon-induced tunneling model written by
Eq.~\ref{eq:T1}. Inset of the graph shows $T_1$ versus temperature
in a linear scale.}
\end{figure}
The $T_1$ for both the N-V and N centers was measured from room
temperature to 40 K. Below 40 K where the $T_1$ is longer than 10
seconds, an accurate measurement proved impractical as the drift
of the superconducting magnet ($\sim 5$ ppm/hour) becomes
nontrivial on the timescale of the measurement. The $T_1$ is
obtained by fitting a decay exponential to the recovery rate of
the echo area $y_0-ae^{-T/T_1}$. As shown in the inset of Fig.~2,
the $T_1$ of both centers increases significantly as the
temperature is reduced. For the N-V center, $T_1$ changes from
$7.7 \pm 0.4$ ms to $3.8 \pm 0.5$ s. For the N center, $T_1$
increases from $1.4 \pm 0.01$ ms to $8.3 \pm 4.7$ s. To evaluate
the temperature dependence of the N center, we applied a
spin-orbit phonon-induced tunneling model which is independent of
the strength of a magnetic field~\cite{reynhardt98}. The
temperature dependence is given by the following,
\begin{eqnarray}\label{eq:T1}
\frac{1}{T_1}=AT+BT^5,
\end{eqnarray}
where A and B are parameters related to Jahn-Teller energy and
electron-phonon interaction~\cite{reynhardt98}. From the fit, we
found $A = 8.0 \times 10^{-3}$ and $B = 3.5 \times 10^{-10}$ which
are in good agreement with the values in Ref.~\cite{reynhardt98},
and confirm a largely field-independent $T_1$ relaxation. The
temperature dependence of the N-V center also shows similar
behavior. The $T_1$ relaxation mechanism for the N-V center is
beyond the scope of this paper~\cite{takahashi08}.

We also investigated the temperature dependence of the spin
coherence time $T_2$ for the N-V center using a Hahn echo sequence
where the width of the pulses (typically 500-700 ns) was tuned to
maximize the echo size. Fig.~3(a) shows the decay of echo area at
room temperature and at $T=1.28\pm0.1$ K. These decays, which are
well fit by a single exponential $e^{-2\tau/T_2}$ as shown in
Fig.~3(a), show no evidence of electron-spin echo envelope
modulation (ESEEM) effects from the $^{14}$N hyperfine
coupling~\cite{schweiger}. This is due to the relative long
microwave pulses and the nuclear Zeeman splitting at 8.5 T which
is much larger than the $^{14}$N hyperfine coupling of the N-V
center. Between room temperature and 20 K, we observe almost no
temperature dependence with $T_2~\ll~T_1$, ({\it e.g.} the $T_2 =
6.7 \pm 0.2~\mu$s at room temperature and $T_2 = 8.3 \pm 0.7~\mu$s
at 20 K). This verifies that the mechanism which determines $T_2$
is different from that of $T_1$. Below the Zeeman energy (11.5 K),
$T_2$ increases drastically as shown in the inset of Fig.~3(b).
\begin{figure}
\includegraphics[width=85 mm]{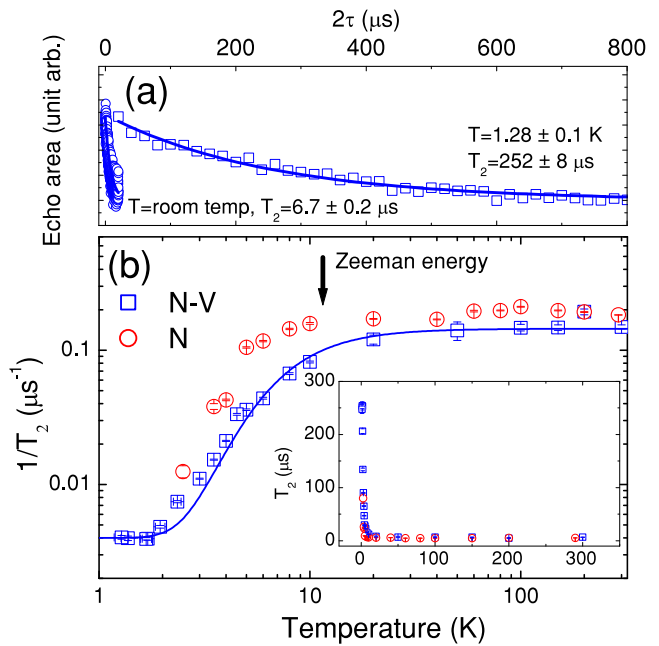}
\caption{\label{fig:invT2G} (a) Echo area of the N-V center as a
function of delay $2\tau$ measured at room temperature and $T=1.28
\pm 0.1$ K. Solid line shows the best fit by the single
exponential. (b)$1/T_2$ for the N-V and N centers versus
temperature. The scale of the main graph is log-log. Solid lines
are the best fit using Eq.~\ref{eq:T2}. The arrow shows the Zeeman
energy of 11.5 K. The inset shows $T_2$ versus temperature in
linear scale which shows a dramatic increase of $T_2$ below the
Zeeman energy.}
\end{figure}
By lowering the temperature further, $T_2$ increases up to $\sim$
250 $\mu$s at 1.7 K and doesn't show noticeable increase below 1.7
K.

At high magnetic field, where single spin flips are suppressed,
the fluctuations in the bath are mainly caused by
energy-conserving flip-flop processes of the N spins. The spin
flip-flop rate in the bath is proportional to the number of pairs
with opposite spin and thus it strongly depends on the spin bath
polarization~\cite{kutter95}. At 240 GHz and 2 K, the N spin bath
polarization is 99.4 $\%$ which almost eliminates the spin
flip-flop process. This experiment therefore verifies that the
dominant decoherence mechanism of the N-V center in type-Ib
diamond is the spin-flop process of the N spin bath. Using the
partition function for the Zeeman term of the N spins,
$Z=\sum_{S=-1/2}^{1/2}e^{-\beta \mu_B g^N B_0 S}$ where
$\beta=1/(k_B T)$ and $k_B$ is Boltzmann constant, the flip-flop
rate is modelled by the following equation~\cite{kutter95},
\begin{eqnarray}\label{eq:T2}
\frac{1}{T_2}\equiv
C P_{m_S=-1/2}P_{m_S=1/2}+\Gamma_{res}\nonumber\\
=\frac{C}{(1+e^{T_{Ze}/T})(1+e^{-T_{Ze}/T})}+\Gamma_{res},
\end{eqnarray}
where C is a temperature independent parameter, $T_{Ze}$ is the
temperature corresponding to Zeeman energy and $\Gamma_{res}$ is a
residual relaxation rate. We fit the $T_2$ data for the N-V center
using the equation above. The fit was performed with the fixed
$\Gamma_{res}=0.004$ ($\mu$s$^{-1}$) corresponding to 250 $\mu$s.
This model fit the data well as shown in the log scale plot of
Fig.~3(b). $T_{Ze} =$ $14.7 \pm 0.4$ K obtained from the fit is in
reasonable agreement with the actual Zeeman energy of 11.5 K. The
result thus confirms the decoherence mechanism of the N spin bath
fluctuation.

The observation of the saturation of $T_2 \sim$ 250 $\mu$s also
indicates complete quenching of the N spin bath fluctuation and a
second decoherence source in this system. From previous
studies~\cite{gaebel06, childress06}, the most probable second
source is a coupling to the $^{13}$C nuclear spin bath. In fact,
$T_2~\sim$ 250 $\mu$s agrees with an estimated decoherence time of
$^{13}$C spin bath fluctuations~\cite{gaebel06}.

Finally we investigate temperature dependence of $T_2$ for the N
center at 240 GHz. No temperature dependence of $T_2$ was observed
in a previous pulsed EPR study at 9.6 GHz~\cite{reynhardt98}. We
measured the $\vert m_S = -1/2, m_I = 1 \rangle \leftrightarrow
\vert 1/2, 1 \rangle$ transition shown in Fig.~1(b) which can
excite only 1/12 of the N center population while it is assumed
that all N spins in this transition are on
resonance~\cite{smith59}. The temperature dependence of $T_2$
therefore shows the relationship between 1/12 of the N center and
11/12 of the N spin bath fluctuation. Similar to the N-V center,
we found slight change between room temperature and 20 K, {\it
i.e.} $T_2 = 5.455 \pm 0.005~\mu$s at room temperature and $T_2 =
5.83 \pm 0.04~\mu$s at 20 K, and then a significant increase below
the Zeeman energy. Eventually, $T_2$ becomes 80 $\pm$ 9 $\mu$s at
2.5 K. As shown in Fig.~3(b), the temperature dependence of $T_2$
is similar to that of the N-V center. These facts support strongly
that the decoherence mechanism of the N center is also the N spin
bath fluctuation.

%
%
In conclusion, we presented the temperature dependence of the spin
relaxation times $T_1$ and $T_2$ of the N-V and N centers in
diamond. The temperature dependence of $T_2$ confirms that the
primary decoherence mechanism in type-Ib diamond is the N spin
bath fluctuation. We have demonstrated that we can strongly
polarize the N spin bath and quench its decoherence at 8 T and 240
GHz. We observed that $T_2$ of the N-V center saturates $\sim$ 250
$\mu$s below 2 K which indicates a secondary decoherence mechanism
and is in good agreement with an estimated coherence time
dominated by $^{13}$C nuclear spin fluctuations.

%
%
This work was supported by research grants; NSF and W. M. Keck
foundation (M.S.S. and S.T.), FOM and NWO (R.H.) and AFOSR
(D.D.A.). S.T. thanks the NHMFL EMR program for travel support.


\begin{thebibliography}{31}
\expandafter\ifx\csname
natexlab\endcsname\relax\def\natexlab#1{#1}\fi
\expandafter\ifx\csname bibnamefont\endcsname\relax
  \def\bibnamefont#1{#1}\fi
\expandafter\ifx\csname bibfnamefont\endcsname\relax
  \def\bibfnamefont#1{#1}\fi
\expandafter\ifx\csname citenamefont\endcsname\relax
  \def\citenamefont#1{#1}\fi
\expandafter\ifx\csname url\endcsname\relax
  \def\url#1{\texttt{#1}}\fi
\expandafter\ifx\csname
urlprefix\endcsname\relax\def\urlprefix{URL }\fi
\providecommand{\bibinfo}[2]{#2}
\providecommand{\eprint}[2][]{\url{#2}}

\bibitem[{\citenamefont{Awschalom and Flatt\'{e}}(2007)}]{awschalom07}
\bibinfo{author}{\bibfnamefont{D.~D.} \bibnamefont{Awschalom}}
  \bibnamefont{and} \bibinfo{author}{\bibfnamefont{M.~E.}
  \bibnamefont{Flatt\'{e}}}, \bibinfo{journal}{Nature Phys.}
  \textbf{\bibinfo{volume}{3}}, \bibinfo{pages}{153} (\bibinfo{year}{2007}).

\bibitem[{\citenamefont{Hanson et~al.}(2007)}]{hanson07}
\bibinfo{author}{\bibfnamefont{R.}~\bibnamefont{Hanson}} \bibnamefont{et~al.},
  \bibinfo{journal}{Rev. Mod. Phys.} \textbf{\bibinfo{volume}{79}},
  \bibinfo{pages}{1217} (\bibinfo{year}{2007}).

\bibitem[{\citenamefont{Khaetskii et~al.}(2002)\citenamefont{Khaetskii, Loss,
  and Glazman}}]{khaetskii02}
\bibinfo{author}{\bibfnamefont{A.~V.} \bibnamefont{Khaetskii}},
  \bibinfo{author}{\bibfnamefont{D.}~\bibnamefont{Loss}}, \bibnamefont{and}
  \bibinfo{author}{\bibfnamefont{L.}~\bibnamefont{Glazman}},
  \bibinfo{journal}{Phys. Rev. Lett.} \textbf{\bibinfo{volume}{88}},
  \bibinfo{pages}{186802} (\bibinfo{year}{2002}).

\bibitem[{\citenamefont{Merkulov et~al.}(2002)\citenamefont{Merkulov, Efros,
  and Rosen}}]{merkulov02}
\bibinfo{author}{\bibfnamefont{I.~A.} \bibnamefont{Merkulov}},
  \bibinfo{author}{\bibfnamefont{Al.~L.} \bibnamefont{Efros}}, \bibnamefont{and}
  \bibinfo{author}{\bibfnamefont{M.}~\bibnamefont{Rosen}},
  \bibinfo{journal}{Phys. Rev. B} \textbf{\bibinfo{volume}{65}},
  \bibinfo{pages}{205309} (\bibinfo{year}{2002}).

\bibitem[{\citenamefont{de~Sousa and Sarma}(2003)}]{desousa03}
\bibinfo{author}{\bibfnamefont{R.}~\bibnamefont{de~Sousa}} \bibnamefont{and}
  \bibinfo{author}{\bibfnamefont{S.} \bibnamefont{Das~Sarma}},
  \bibinfo{journal}{Phys. Rev. B} \textbf{\bibinfo{volume}{68}},
  \bibinfo{pages}{115322} (\bibinfo{year}{2003}).

\bibitem[{\citenamefont{Dobrovitski et~al.}(2003)\citenamefont{Dobrovitski,
  Raedt, Katsnelson, and Harmon}}]{dobrovitski03}
\bibinfo{author}{\bibfnamefont{V.~V.} \bibnamefont{Dobrovitski}},
  \bibinfo{author}{\bibfnamefont{H.~A.} \bibnamefont{De~Raedt}},
  \bibinfo{author}{\bibfnamefont{M.~I.} \bibnamefont{Katsnelson}},
  \bibnamefont{and} \bibinfo{author}{\bibfnamefont{B.~N.}
  \bibnamefont{Harmon}}, \bibinfo{journal}{Phys. Rev. Lett.}
  \textbf{\bibinfo{volume}{90}}, \bibinfo{pages}{210401}
  (\bibinfo{year}{2003}).

\bibitem[{\citenamefont{Cucchietti et~al.}(2005)\citenamefont{Cucchietti, Paz,
  and Zurek}}]{cucchietti05}
\bibinfo{author}{\bibfnamefont{F.~M.} \bibnamefont{Cucchietti}},
  \bibinfo{author}{\bibfnamefont{J.~P.} \bibnamefont{Paz}}, \bibnamefont{and}
  \bibinfo{author}{\bibfnamefont{W.~H.} \bibnamefont{Zurek}},
  \bibinfo{journal}{Phys. Rev. A} \textbf{\bibinfo{volume}{72}},
  \bibinfo{pages}{052113} (\bibinfo{year}{2005}).

\bibitem[{\citenamefont{Taylor and Lukin}(2006)}]{taylor06}
\bibinfo{author}{\bibfnamefont{J.~M.} \bibnamefont{Taylor}} \bibnamefont{and}
  \bibinfo{author}{\bibfnamefont{M.~D.} \bibnamefont{Lukin}},
  \bibinfo{journal}{Quant. Info. Proc.} \textbf{\bibinfo{volume}{5}},
  \bibinfo{pages}{503} (\bibinfo{year}{2006}).

\bibitem[{\citenamefont{Yao et~al.}(2007)\citenamefont{Yao, Liu, and
  Sham}}]{yao07}
\bibinfo{author}{\bibfnamefont{W.}~\bibnamefont{Yao}},
  \bibinfo{author}{\bibfnamefont{R.-B.} \bibnamefont{Liu}}, \bibnamefont{and}
  \bibinfo{author}{\bibfnamefont{L.~J.} \bibnamefont{Sham}},
  \bibinfo{journal}{Phys. Rev. Lett.} \textbf{\bibinfo{volume}{98}},
  \bibinfo{pages}{077602} (\bibinfo{year}{2007}).

\bibitem[{\citenamefont{Stepanenko et~al.}(2006)\citenamefont{Stepanenko,
  Burkard, Giedke, and Imamoglu}}]{stepanenko06}
\bibinfo{author}{\bibfnamefont{D.}~\bibnamefont{Stepanenko}},
  \bibinfo{author}{\bibfnamefont{G.}~\bibnamefont{Burkard}},
  \bibinfo{author}{\bibfnamefont{G.}~\bibnamefont{Giedke}}, \bibnamefont{and}
  \bibinfo{author}{\bibfnamefont{A.}~\bibnamefont{Imamoglu}},
  \bibinfo{journal}{Phys. Rev. Lett.} \textbf{\bibinfo{volume}{96}},
  \bibinfo{pages}{136401} (\bibinfo{year}{2006}).

\bibitem[{\citenamefont{Klauser et~al.}(2006)\citenamefont{Klauser, Coish, and
  Loss}}]{klauser06}
\bibinfo{author}{\bibfnamefont{D.}~\bibnamefont{Klauser}},
  \bibinfo{author}{\bibfnamefont{W.~A.} \bibnamefont{Coish}}, \bibnamefont{and}
  \bibinfo{author}{\bibfnamefont{D.}~\bibnamefont{Loss}},
  \bibinfo{journal}{Phys. Rev. B} \textbf{\bibinfo{volume}{73}},
  \bibinfo{pages}{205302} (\bibinfo{year}{2006}).

\bibitem[{\citenamefont{Bracker et~al.}(2005)}]{bracker05}
\bibinfo{author}{\bibfnamefont{A.~S.} \bibnamefont{Bracker}}
  \bibnamefont{et~al.}, \bibinfo{journal}{Phys. Rev. Lett.}
  \textbf{\bibinfo{volume}{94}}, \bibinfo{pages}{047402}
  (\bibinfo{year}{2005}).

\bibitem[{\citenamefont{Baugh et~al.}(2007)\citenamefont{Baugh, Kitamura, Ono,
  and Tarucha}}]{baugh07}
\bibinfo{author}{\bibfnamefont{J.}~\bibnamefont{Baugh}},
  \bibinfo{author}{\bibfnamefont{Y.}~\bibnamefont{Kitamura}},
  \bibinfo{author}{\bibfnamefont{K.}~\bibnamefont{Ono}}, \bibnamefont{and}
  \bibinfo{author}{\bibfnamefont{S.}~\bibnamefont{Tarucha}},
  \bibinfo{journal}{Phys. Rev. Lett.} \textbf{\bibinfo{volume}{99}},
  \bibinfo{pages}{096804} (\bibinfo{year}{2007}).

\bibitem[{\citenamefont{Coish and Loss}(2004)}]{coish04}
\bibinfo{author}{\bibfnamefont{W.~A.} \bibnamefont{Coish}} \bibnamefont{and}
  \bibinfo{author}{\bibfnamefont{D.}~\bibnamefont{Loss}},
  \bibinfo{journal}{Phys. Rev. B} \textbf{\bibinfo{volume}{70}},
  \bibinfo{pages}{195340} (\bibinfo{year}{2004}).

\bibitem[{\citenamefont{Kennedy et~al.}(2003)}]{kennedy03}
\bibinfo{author}{\bibfnamefont{T.~A.} \bibnamefont{Kennedy}}
  \bibnamefont{et~al.}, \bibinfo{journal}{Appl. Phys. Lett.}
  \textbf{\bibinfo{volume}{83}}, \bibinfo{pages}{4190} (\bibinfo{year}{2003}).

\bibitem[{\citenamefont{Gaebel et~al.}(2006)}]{gaebel06}
\bibinfo{author}{\bibfnamefont{T.}~\bibnamefont{Gaebel}} \bibnamefont{et~al.},
  \bibinfo{journal}{Nature Phys.} \textbf{\bibinfo{volume}{2}},
  \bibinfo{pages}{408} (\bibinfo{year}{2006}).

\bibitem[{\citenamefont{Hanson et~al.}(2006{\natexlab{a}})\citenamefont{Hanson,
  Mendoza, Epstein, and Awschalom}}]{hanson06prl}
\bibinfo{author}{\bibfnamefont{R.}~\bibnamefont{Hanson}},
  \bibinfo{author}{\bibfnamefont{F.~M.} \bibnamefont{Mendoza}},
  \bibinfo{author}{\bibfnamefont{R.~J.} \bibnamefont{Epstein}},
  \bibnamefont{and} \bibinfo{author}{\bibfnamefont{D.~D.}
  \bibnamefont{Awschalom}}, \bibinfo{journal}{Phys. Rev. Lett.}
  \textbf{\bibinfo{volume}{97}}, \bibinfo{pages}{087601}
  (\bibinfo{year}{2006}{\natexlab{a}}).

\bibitem[{\citenamefont{Hanson et~al.}(2008)\citenamefont{Hanson, Dobrovitski,
  Feiguin, Gywat, and Awschalom}}]{hanson08}
\bibinfo{author}{\bibfnamefont{R.}~\bibnamefont{Hanson}},
  \bibinfo{author}{\bibfnamefont{V.~V.} \bibnamefont{Dobrovitski}},
  \bibinfo{author}{\bibfnamefont{A.~E.} \bibnamefont{Feiguin}},
  \bibinfo{author}{\bibfnamefont{O.}~\bibnamefont{Gywat}}, \bibnamefont{and}
  \bibinfo{author}{\bibfnamefont{D.~D.} \bibnamefont{Awschalom}},
  \bibinfo{journal}{Science}, \bibinfo{pages}{March 13}
  (\bibinfo{year}{2008}), \bibinfo{note}{(10.1126/science.1155400)}.

\bibitem[{\citenamefont{Childress et~al.}(2006)}]{childress06}
\bibinfo{author}{\bibfnamefont{L.}~\bibnamefont{Childress}}
  \bibnamefont{et~al.}, \bibinfo{journal}{Science}
  \textbf{\bibinfo{volume}{314}}, \bibinfo{pages}{281} (\bibinfo{year}{2006}).

\bibitem[{\citenamefont{Dutt et~al.}(2007)}]{dutt07}
\bibinfo{author}{\bibfnamefont{M.~V.} \bibnamefont{Gurudev~Dutt}}
  \bibnamefont{et~al.}, \bibinfo{journal}{Science}
  \textbf{\bibinfo{volume}{316}}, \bibinfo{pages}{1312} (\bibinfo{year}{2007}).

\bibitem[{\citenamefont{Hanson et~al.}(2006{\natexlab{b}})\citenamefont{Hanson,
  Gywat, and Awschalom}}]{hanson06}
\bibinfo{author}{\bibfnamefont{R.}~\bibnamefont{Hanson}},
  \bibinfo{author}{\bibfnamefont{O.}~\bibnamefont{Gywat}}, \bibnamefont{and}
  \bibinfo{author}{\bibfnamefont{D.~D.} \bibnamefont{Awschalom}},
  \bibinfo{journal}{Phys. Rev. B} \textbf{\bibinfo{volume}{74}},
  \bibinfo{pages}{161203R} (\bibinfo{year}{2006}{\natexlab{b}}).

\bibitem[{\citenamefont{Epstein et~al.}(2005)\citenamefont{Epstein, Mendoza,
  Kato, and Awschalom}}]{epstein05}
\bibinfo{author}{\bibfnamefont{R.~J.} \bibnamefont{Epstein}},
  \bibinfo{author}{\bibfnamefont{F.~M.} \bibnamefont{Mendoza}},
  \bibinfo{author}{\bibfnamefont{Y.~K.} \bibnamefont{Kato}}, \bibnamefont{and}
  \bibinfo{author}{\bibfnamefont{D.~D.} \bibnamefont{Awschalom}},
  \bibinfo{journal}{Nature Phys.} \textbf{\bibinfo{volume}{1}},
  \bibinfo{pages}{94} (\bibinfo{year}{2005}).

\bibitem[{\citenamefont{Loubser and vanWyk}(1978)}]{loubser78}
\bibinfo{author}{\bibfnamefont{J.~H.~N.} \bibnamefont{Loubser}}
  \bibnamefont{and} \bibinfo{author}{\bibfnamefont{J.~A.}
  \bibnamefont{vanWyk}}, \bibinfo{journal}{Rep. Prog. Phys.}
  \textbf{\bibinfo{volume}{41}}, \bibinfo{pages}{1201} (\bibinfo{year}{1978}).

\bibitem[{\citenamefont{Smith et~al.}(1959)\citenamefont{Smith, Sorokin,
  Gelles, and Lasher}}]{smith59}
\bibinfo{author}{\bibfnamefont{W.~V.} \bibnamefont{Smith}},
  \bibinfo{author}{\bibfnamefont{P.~P.} \bibnamefont{Sorokin}},
  \bibinfo{author}{\bibfnamefont{I.~L.} \bibnamefont{Gelles}},
  \bibnamefont{and} \bibinfo{author}{\bibfnamefont{G.~J.}
  \bibnamefont{Lasher}}, \bibinfo{journal}{Phys. Rev.}
  \textbf{\bibinfo{volume}{115}}, \bibinfo{pages}{1546} (\bibinfo{year}{1959}).

\bibitem[{\citenamefont{van Tol et~al.}(2005)\citenamefont{van Tol, Brunel, and
  Wylde}}]{vantol05}
\bibinfo{author}{\bibfnamefont{J.}~\bibnamefont{van Tol}},
  \bibinfo{author}{\bibfnamefont{L.-C.} \bibnamefont{Brunel}},
  \bibnamefont{and} \bibinfo{author}{\bibfnamefont{R.~J.} \bibnamefont{Wylde}},
  \bibinfo{journal}{Rev. Sci. Instrum.} \textbf{\bibinfo{volume}{76}},
  \bibinfo{pages}{074101} (\bibinfo{year}{2005}).

\bibitem[{\citenamefont{Morley et~al.}(2008)\citenamefont{Morley, Brunel, and
  van Tol}}]{morley08}
\bibinfo{author}{\bibfnamefont{G.~W.} \bibnamefont{Morley}},
  \bibinfo{author}{\bibfnamefont{L.-C.} \bibnamefont{Brunel}},
  \bibnamefont{and} \bibinfo{author}{\bibfnamefont{J.}~\bibnamefont{van Tol}},
  \bibinfo{note}{arXiv:0803.3054}.

\bibitem[{\citenamefont{Zhang et~al.}(1994)}]{zhang94}
\bibinfo{author}{\bibfnamefont{S.}~\bibnamefont{Zhang}} \bibnamefont{et~al.},
  \bibinfo{journal}{Phys. Rev. B} \textbf{\bibinfo{volume}{49}},
  \bibinfo{pages}{15392} (\bibinfo{year}{1994}).

\bibitem[{\citenamefont{Schweiger and Jeschke}(2001)}]{schweiger}
\bibinfo{author}{\bibfnamefont{A.}~\bibnamefont{Schweiger}} \bibnamefont{and}
  \bibinfo{author}{\bibfnamefont{G.}~\bibnamefont{Jeschke}}, in
  \emph{\bibinfo{booktitle}{Principles of Pulse Electron Paramagnetic
  Resonance}} (\bibinfo{publisher}{Oxford Univeristy Press},
  \bibinfo{address}{New York}, \bibinfo{year}{2001}).

\bibitem[{\citenamefont{Reynhardt et~al.}(1998)\citenamefont{Reynhardt, High,
  and vanWyk}}]{reynhardt98}
\bibinfo{author}{\bibfnamefont{E.~C.} \bibnamefont{Reynhardt}},
  \bibinfo{author}{\bibfnamefont{G.~L.} \bibnamefont{High}}, \bibnamefont{and}
  \bibinfo{author}{\bibfnamefont{J.~A.} \bibnamefont{vanWyk}},
  \bibinfo{journal}{J. Chem. Phys.} \textbf{\bibinfo{volume}{109}},
  \bibinfo{pages}{8471} (\bibinfo{year}{1998}).

\bibitem[{\citenamefont{Takahashi et~al.}(2008)}]{takahashi08}
\bibinfo{author}{\bibfnamefont{S.}~\bibnamefont{Takahashi}}
  \bibnamefont{et~al.}, (\bibinfo{note}{to be published}).

\bibitem[{\citenamefont{Kutter et~al.}(1995)}]{kutter95}
\bibinfo{author}{\bibfnamefont{C.}~\bibnamefont{Kutter}} \bibnamefont{et~al.},
  \bibinfo{journal}{Phys. Rev. Lett.} \textbf{\bibinfo{volume}{74}},
  \bibinfo{pages}{2925} (\bibinfo{year}{1995}).

\end{thebibliography}
\end{document}